\documentclass{article}
\usepackage[utf8]{inputenc}

\pdfoutput=1
\usepackage{amssymb}
\usepackage{amsmath}
\usepackage[dvips]{graphicx}
\usepackage{setspace}
\usepackage{slashed}
\usepackage{mathtools}
\usepackage{amsfonts}
\usepackage{fancyhdr}
\usepackage{xcolor}
\usepackage{graphicx}
\usepackage{rotating}
\usepackage{comment}
\usepackage{color}
\usepackage{subcaption}
\usepackage[percent]{overpic}
\usepackage{cite}
\usepackage{braket}
\usepackage{moresize}
\definecolor{darkgreen}{rgb}{0,0.5,0}
\definecolor{darkblue}{rgb}{0,0,0.6}
\definecolor{purple}{rgb}{0.4,.2,0.7}
\newcommand{\p}{\partial}

\newcommand{\be}{\begin{equation}}
\newcommand{\ee}{\end{equation}}

\usepackage[colorlinks=true,citecolor=darkgreen,linkcolor=black,urlcolor=purple]{hyperref}

\usepackage{pdfsync}

\makeatletter
\newcommand*{\defeq}{\mathrel{\rlap{%
                     \raisebox{0.3ex}{$\m@th\cdot$}}%
                     \raisebox{-0.3ex}{$\m@th\cdot$}}%
                     =} 
\makeatother

\DeclareMathOperator{\Tr}{Tr}
\def\be{\begin{eqnarray}}
\def\ee{\end{eqnarray}}

\newcommand{\tr}{\textrm{Tr}\,}

\newcommand{\bea}{\begin{eqnarray}}
\newcommand{\eea}{\end{eqnarray}}
\def\ben{\begin{equation}}
\def\een{\end{equation}}

    \let\p=\phi \let\r=v

\def\be{\begin{equation}}
\def\ee{\end{equation}}
\def\ba{\begin{eqnarray}}
\def\ea{\end{eqnarray}}

\def\bal#1\eal{\begin{align}#1\end{align}}
\def\bs#1\es{\begin{split}#1\end{split}}

\renewcommand{\p}{\partial}

\allowdisplaybreaks  

\numberwithin{equation}{section}

\def\p{{\phi}}

\def\be{\begin{equation}}
\def\ee{\end{equation}}
\def\ba{\begin{eqnarray}}
\def\ea{\end{eqnarray}}
\def\bal#1\eal{\begin{align}#1\end{align}}

\def\r{\rightarrow}

\def\r{\right}

\usepackage{tikz}
\usetikzlibrary{positioning,arrows}
\usetikzlibrary{decorations.pathmorphing}
\usetikzlibrary{decorations.markings}
\tikzset{
particle/.style={postaction={decorate}},
graviton/.style={decorate, decoration={snake, amplitude=0.8 mm, segment length=1.5 mm, pre length=0.8 mm, post length=0.8 mm}},
photon/.style={
        decoration={complete sines, amplitude=0.15cm, segment length=0.2cm},
        decorate    
    },
gluon/.style={
        decoration={coil, aspect=0.75, mirror, segment length=1.5mm},
        decorate
    }
}
 
\def \be {\begin{equation}}
\def \ee {\end{equation}}

\renewcommand{\p}{\partial}

\newcommand{\tI}{\widetilde{I}}

\newcommand{\M}{\mathcal{M}}
\newcommand{\B}{\mathcal{B}}

\usepackage{framed}

\begin{document}
\onehalfspacing

\begin{center}

~
\vskip5mm

{\LARGE  {
Black hole wavefunctions and microcanonical states
}}

\vskip7mm

Wan Zhen Chua and Thomas Hartman

\vskip5mm

{\it Department of Physics, Cornell University, Ithaca, New York, USA
}

\vskip5mm

\end{center}

\vspace{2mm}

\begin{abstract}
\noindent
We consider the problem of defining a microcanonical thermofield double state at fixed energy and angular momentum from the gravitational path integral. A semiclassical approximation to this state is obtained by imposing a mixed boundary condition on an initial time surface. We analyze the corresponding boundary value problem and gravitational action. The overlap of this state with the canonical thermofield double state, which is interpreted as the Hartle-Hawking wavefunction of an eternal black hole in a mini-superspace approximation, is calculated semiclassically. The relevant saddlepoint is a higher-dimensional, rotating generalization of the wedge geometry that has been studied in two-dimensional gravity.

 \end{abstract}

\pagebreak
\pagestyle{plain}

\setcounter{tocdepth}{2}
{}
\vfill
\tableofcontents


\date{}

\section{Introduction}

The eternal Schwarzschild black hole in anti-de Sitter space is holographically dual to the thermofield double (TFD) state \cite{Maldacena:2001kr},
\begin{align}\label{introTFD}
    |\Psi_{\beta/2}\rangle = \sum_{n} e^{-\beta E_n/2}|n^*\rangle|n\rangle
\end{align}
where $n$ labels energy eigenstates and $*$ is the CPT conjugate. 
In this paper, we study the wavefunction for this state on the gravity side. In principle, the wavefunction depends on an infinite amount of data, but we will project onto states $|E,J_I\rangle$ labeled only by energy and angular momenta. These states are defined by a boundary-value problem in the bulk, with the data $(E, J_I)$ specifying the boundary conditions on a spatial slice $\Sigma$. The prescription is a mixed boundary condition that fixes some components of the induced metric and some components of the extrinsic curvature on $\Sigma$.

The calculation  of the wavefunction is performed semiclassically, with the result
\begin{align}\label{introWF}
    \Psi_{\beta/2}(E,J_I) = \langle E,J_I| \Psi_{\beta/2}\rangle \approx e^{S(E,J_I)/2 - \beta E/2}
\end{align}
where $S$ is the black hole entropy.
This is the wavefunction of a non-rotating black hole at temperature $\beta$, which depends independently on the parameters $E, J_I$, and $\beta$. 
The eternal black hole is the saddlepoint that appears in the thermal partition function $Z(\beta) = \langle \Psi_{\beta/2} | \Psi_{\beta/2}\rangle$. The wavefunction \eqref{introWF} contains additional information; for example, it can also be used to calculate the partition function of a rotating black hole, using $Z(\beta,\theta^I) = \langle \Psi_{\beta/2} | e^{i \theta^I J_I} | \Psi_{\beta/2}\rangle$. The derivation of \eqref{introWF} does not rely on the AdS asymptotics in any essential way so it also holds in asymptotically flat space, with the usual caveat that the canonical ensemble is unstable. 

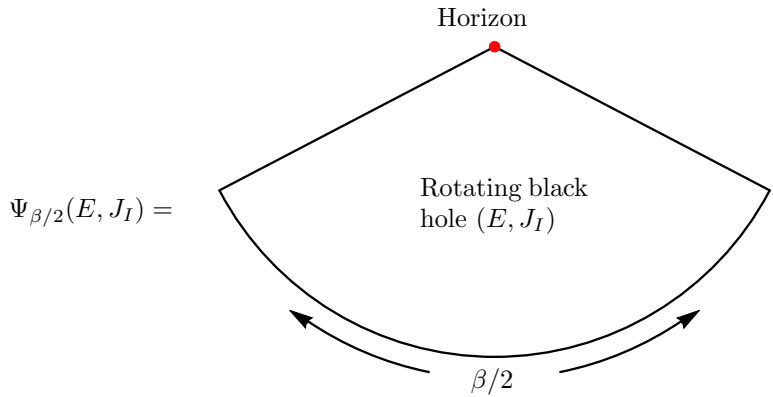
\begin{figure}
	\begin{center}\hspace{1cm}
		\begin{overpic}[scale=0.6]{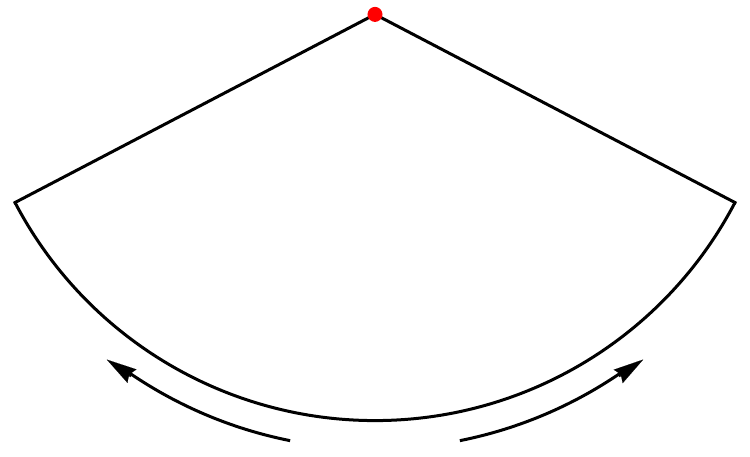}
	\put(30,-1){\parbox{0.2\linewidth}{
		\begin{equation*}
			\beta/2
\end{equation*}}}
	\put(-40,30){\parbox{0.2\linewidth}{
		\begin{equation*}
		\Psi_{\beta/2}(E,J_I)=
\end{equation*}}}
	\put(40,63){\parbox{0.2\linewidth}{Horizon}}
		\put(37,30){\parbox{0.15\linewidth}{Rotating black hole $(E,J_I)$}}
		\end{overpic}  
	\end{center}
	\caption{\small The Hartle-Hawking state $|\Psi_{\beta/2}\rangle$ is prepared by a Euclidean gravitational path integral in which the boundary condition is a strip of length $\beta/2$ at the asymptotic AdS boundary. The state $|E,J_I\rangle$ is defined by a boundary condition on a slice $\Sigma$ through the bulk. The wavefunction is the overlap $\Psi_{\beta/2} (E, J_I) = \langle E, J_I | \Psi_{\beta/2} \rangle \approx e^{-\tI}$, where $\tI$ is the action of the saddle. The saddle satisfying these boundary conditions is a wedge cut from the rotating eternal black hole.\label{fig:wavefunction-pacman}}
\end{figure}

Microscopically, the state $|E, J_I\rangle$ is interpreted as a semiclassical approximation to the microcanonical thermofield double,
\begin{align}\label{introEJ}
|E, J_I\rangle &=  e^{-S(E,J_I)/2} \sum_{{\cal H}(E,J_I)} |n^*\rangle |n\rangle \ ,
\end{align}
where the sum is over states in a microcanonical window (the precise definition of this window does not affect the leading semiclassical answers and is not addressed). From a dual CFT point of view, the overlap \eqref{introWF} follows trivially from \eqref{introTFD} and \eqref{introEJ}. The new ingredient in our calculation is to carefully setup the gravitational boundary-value problem that defines the state $| E,J_I\rangle$ and to reproduce \eqref{introWF} from the bulk. 
The saddlepoint responsible for \eqref{introWF} is the wedge (or `pacman') geometry illustrated in figure \ref{fig:wavefunction-pacman}. The corner in this geometry is not part of the boundary condition; it occurs dynamically on the saddle. The boundary conditions allow corners to occur only at extremal surfaces.

In two-dimensional JT gravity, the 2d version of this geometry was used to calculate the Hartle-Hawking wavefunction semiclassically by Harlow and Jafferis \cite{Harlow:2018tqv} and exactly by Yang \cite{Yang:2018gdb}. It has also been generalized to include bulk matter coupled to JT gravity \cite{Saad:2019pqd}. These results have proved useful in calculating higher topology contributions to the gravitational path integral. One of our primary motivations for lifting the calculation to higher dimensions is to study higher topologies in higher dimensions (see e.g. 
\cite{Maldacena:2004rf,Marolf:2021kjc,Penington:2019kki,Almheiri:2019qdq,Cotler:2020ugk,Chandra:2022bqq,Chandra:2022fwi,Chandra:2023dgq,Chandra:2023rhx,Collier:2023fwi,Balasubramanian:2022gmo,Bah:2022uyz}).  An important new ingredient in higher dimensions is rotation; the wavefunction involves a novel corner term in the gravitational action that arises when two manifolds are glued with a relative twist, which occurs at a rotating extremal surface.

\subsection{Comparison to the literature}

There is a large literature on the microcanonical gravitational path integral, starting with the work of Brown and York \cite{Brown:1992bq}. We will rely on the Brown-York formalism, but the question we seek to address is different. Brown and York calculate the microcanonical partition function $Z_{\rm micro} = e^{S}$ from the gravitational path integral and as such the relevant saddlepoints have only asymptotic boundaries. In contrast, we are using the path integral to define a microcanonical \textit{state}, so there is an internal spacelike boundary $\Sigma$ as well as the asymptotic boundary. Our fixed-$(E,J_I)$ boundary conditions are imposed at this internal boundary, not at asymptotic infinity.   See section \ref{ss:ejbc} for a more detailed comparison to Brown and York.

 Marolf \cite{Marolf:2018ldl} defined a microcanonical, fixed-energy TFD state from the gravitational path integral by writing it as a superposition of canonical TFD states with complex temperatures. Since the canonical TFD state has a path integral preparation, this defines a microcanonical state as a sum of gravitational path integrals. Semiclassical calculations are nonetheless dominated by a single saddlepoint \cite{Marolf:2018ldl}. Our fixed-energy results agree at leading order, as they must, but there are some differences in the details. Aside from including angular momentum, the main difference is that we fix data on a bulk slice, not just at infinity. This has the disadvantage that our state can only be viewed as a semiclassical approximation to the true microcanonical TFD; the advantages are that the state does not require taking a superposition, and that it can more easily be used as a building block for higher topology saddles cut along bulk slices.

Also closely related is the work of Dong, Harlow, and Marolf \cite{Dong:2018seb} on fixed area states in the gravitational path integral. See also \cite{Dong:2019piw,Dong:2022ilf}.
These papers focus on partition functions --- that is, path integrals with no internal boundaries --- rather than states or wavefunctions, but the two are closely related. As explained in \cite{Dong:2018seb,Dong:2022ilf} the fixed-area version of the canonical TFD is semiclassically almost the same as the microcanonical TFD. (See also \cite{Goel:2020yxl} for a discussion of fixed area states and the microcanonical TFD in the sense of \cite{Marolf:2018ldl}.) Indeed, it is not difficult to calculate the wavefunction \eqref{introWF} using fixed-area methods, but the calculation is somewhat indirect (it involves fixing a defect angle first and then performing a Legendre transform). We therefore find it useful to pursue a more direct approach along the lines of \cite{Harlow:2018tqv} where the fixed-$(E,J_I)$ state is defined explicitly as a boundary value problem on a bulk slice.  

Takayanagi and Tamaoka \cite{Takayanagi:2019tvn} have also considered the role of corner terms in black hole entropy and AdS/BCFT. This involves some similar ingredients, including the gluing rules that we discuss below, but the boundary value problem that we consider is different. Gravitational boundary value problems can have pathologies in perturbation theory (see e.g.~\cite{Witten:2018lgb, Marolf:2022jra}), so it would be interesting to check whether our results can be consistently extended to include fluctuations.

In \cite{Chua}, the Hartle-Hawking state in 3d gravity is related to two copies of the ZZ state in Liouville theory. This allows for an exact calculation of the wavefunction in the space of Ishibashi states, which agrees with our results in the semiclassical limit $G \to 0$.

\subsection{Black hole wavefunctions}

In the rest of this introduction we will define black hole wavefunctions in more detail, and explain how they are related to partition functions.

The Hartle-Hawking (HH) state is defined by the Euclidean path integral \cite{Hartle:1976tp}
\begin{equation}\label{hhformal}
	|\Psi_{\beta/2}\rangle = 
 \vcenter{ \hbox{	\begin{overpic}[scale=0.4]{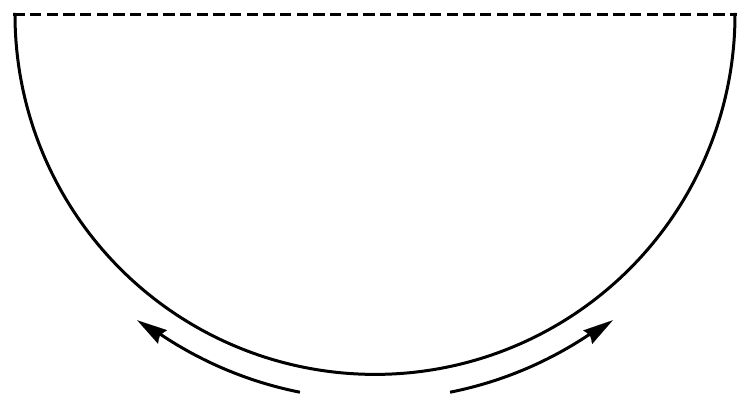}
		\put(20,-3){\parbox{0.2\linewidth}{
				\begin{equation*}
				\beta/2
		\end{equation*}}}
	\end{overpic}}}~.
\end{equation}
This diagram represents the boundary conditions in the path integral. Only the $(r,\tau)$ directions are drawn, with the transverse directions suppressed. The solid semicircle is the asymptotic boundary, where we impose standard asymptotically-AdS boundary conditions, with Euclidean time running over the interval $\tau \in [0, \frac{\beta}{2}]$.  The dashed line represents an open cut on which we must insert the state $\langle \varphi|$ in order to calculate the wavefunction $\langle \varphi | \Psi_{\beta/2}\rangle$. In principle there can be higher topology (or even fully non-geometric) contributions to the quantum gravity path integral, but we define the semiclassical wavefunction by restricting to  geometries with the topology Disk$\times S^{d-1}$.  The state $|\Psi_{\beta/2}\rangle$ is dual to the thermofield double state in two copies of the CFT Hilbert space, $|\Psi_{\rm \beta/2}\rangle \in {\cal H}_{\rm CFT} \times {\cal H}_{\rm CFT}$ \cite{Maldacena:2001kr}. 

We will define a gravitational state of fixed energy $|E\rangle$ and another state of fixed energy and angular momentum $|E, J_I\rangle$ by imposing boundary conditions on a spacelike slice $\Sigma$ through the bulk. In AdS$_D$, with $D=d+1$, the index $I$ runs over the $\lfloor \frac{D-1}{2}\rfloor$ independent angular momenta. Roughly speaking, our boundary conditions require the geometry on $\Sigma$ to be similar to an eternal black hole at energy $E$ and angular momentum $J_I$, but of course we cannot specify both the spatial metric $h_{ij}$ and and the extrinsic curvature $K_{ij}$, since they are conjugate. We will fix the transverse (angular) components of the metric, and the canonical momenta conjugate to the other components. This requires adding boundary terms at $\Sigma$ to the gravitational action in order to have a good variational principle. 

Unlike the Hartle-Hawking state, the path integral definition of the state $|E, J_I\rangle$ does not involve any time evolution. It is a boundary condition on an initial time surface. This surface can be embedded in either a Euclidean or Lorentzian spacetime.\footnote{Our conventions are such that Euclidean black holes with a real metric have real $\theta^I$ and imaginary $J_I$, while Lorentzian black holes with a real metric have imaginary $\theta^I$ and real $J_I$. The canonical ensemble is $Z = \Tr e^{-\beta H + i \theta^I J_I}$. Physical states (e.g.~in the Hilbert space of the dual CFT) have real spin. From a CFT point of view it does not make sense to consider a microcanonical TFD at imaginary $J_I$, but the bulk state $|E, J_I\rangle$ does make sense for imaginary $J_I$, and it is for imaginary $J_I$ that the Euclidean saddles are real. In all of our gravity calculations, $J_I$ can be complex.}

The overlap of $|E, J_I\rangle$ with the Hartle-Hawking state is the wavefunction $\Psi_{\beta/2}(E,J_I)$. This is calculated semiclassically by a gravitational path integral with HH boundary conditions at the asymptotic boundary, and fixed-$(E,J_I)$ boundary conditions at the internal boundary. The saddlepoint is the wedge shown in figure \ref{fig:wavefunction-pacman}. 

We will now discuss some simple applications of the wavefunction, and the relations used to glue multiple wedges together semiclassically. 

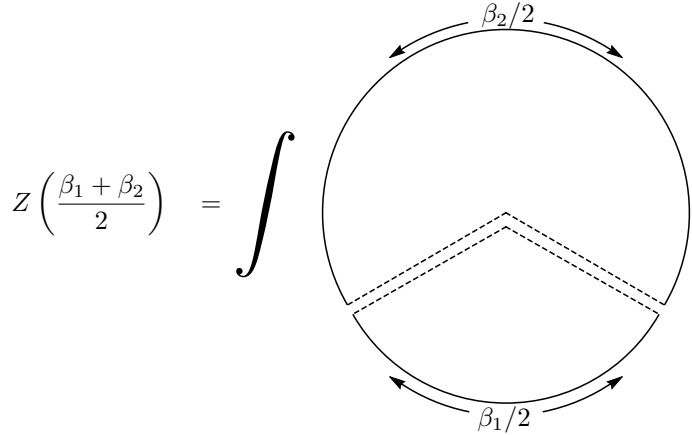
\begin{figure}
	\begin{center}\hspace{2.5cm}
		\begin{overpic}[scale=0.4]{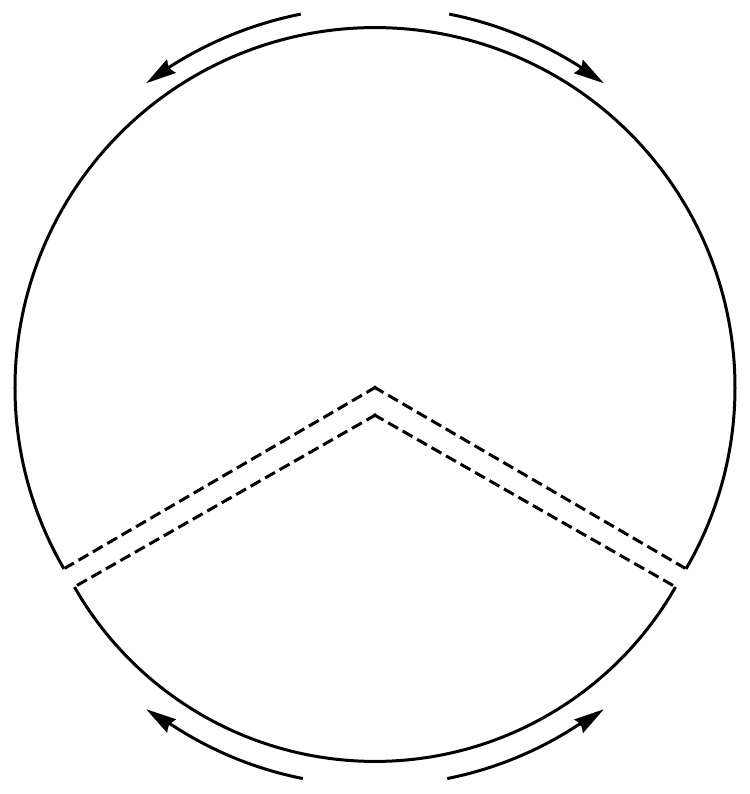}
			\put(19,-2){\parbox{0.2\linewidth}{
					\begin{equation*}
						\beta_1/2
			\end{equation*}}}
					\put(20,98){\parbox{0.2\linewidth}{
				\begin{equation*}
					\beta_2/2
		\end{equation*}}}
					\put(-75,52){\parbox{0.2\linewidth}{
		\begin{equation*}
Z\left(\frac{\beta_1+\beta_2}{2}\right) \quad  = \quad
\end{equation*}}}
					\put(-40,50){\parbox{0.2\linewidth}{
		\begin{equation*}
\text{\Huge  $\int$}
\end{equation*}}}
		\end{overpic}  
	\end{center}
    \caption{\small
	The eternal AdS-Schwarzschild black hole as the inner produce $Z(\frac{\beta_1+\beta_2}{2}) = \langle \Psi_{\beta_2/2}|\Psi_{\beta_1/2}\rangle$.
}
\label{fig:glue2}
\end{figure}

Given the wavefunction \eqref{introWF} we can calculate the thermal partition function by inserting 
\begin{align}\label{incompleteset}
\int dE dJ_I |E, J_I\rangle \langle E, J_I| \ . 
\end{align}
This is not a complete set of states, but it suffices to calculate the gravitational action in situations where the saddle has a slice matching that of a two-sided eternal black hole. The thermal partition function, for example, is
\begin{align}\label{zbeta1}
Z(\beta) = \langle \Psi_{\beta/2} | \Psi_{\beta/2}\rangle \approx \int dE dJ_I e^{S(E, J_I) - \beta E } 
\end{align}
with the integral calculated by a saddlepoint, which manifestly reproduces the usual canonical partition function.
Geometrically, the saddlepoint in this integral is the (entire) eternal black hole. We can also obtain the thermal partition function from the overlap
\begin{align}\label{zbeta2}
Z\left(\frac{\beta_1+\beta_2}{2}\right) = \langle \Psi_{\beta_1/2} | \Psi_{\beta_2/2}\rangle  
\end{align}
This is shown pictorially in figure \ref{fig:glue2}. 

In both \eqref{zbeta1} and \eqref{zbeta2}, the saddlepoint lands at $J_I=0$, because the boundary condition in \eqref{hhformal} does not involve any angular potential. However, the $J_I$ dependence of the wavefunction is also meaningful, because the the non-rotating HH wavefunction can be used to calculate the partition function at finite temperature and angular potential:
\begin{align}
    Z(\beta,\theta^I) = \langle \Psi_{\beta/2} | e^{i \theta^I J_I} | \Psi_{\beta/2}\rangle 
    = \int dM dJ e^{S(E,J_I) - \beta E + i \theta^I J_I}
\end{align}
The insertion of $e^{i\theta^I J_I}$ is realized geometrically by gluing one of the two asymptotic boundaries with a relative twist by $\theta^I$ along the transverse $S^{D-2}$. The corresponding saddlepoint is the Kerr-AdS (or Myers-Perry-AdS) black hole.\footnote{
The reader may wonder why we have only considered the HH state labeled by inverse temperature $\beta$, rather than the more general rotating HH state labeled by $(\beta, \theta^I)$. The reason is that these states, as defined by the path integral, are essentially the same. The only difference is how we label the angular directions at the open cut (at the top-left and top-right corners of the diagram in \eqref{hhformal}). The state is specified not just by this geometry, but also by choosing a marked point (the `origin') on the transverse space at the corners. The marked point on one side can be set to zero; the other is physical. It specifies how to glue the ket to another state $\langle \varphi|$. Inserting the operator $e^{i \theta^I J_I}$ simply moves the marked point.  
For an example, take $D=3$. The HH state $|\Psi_{\beta/2}\rangle$ on the 2d boundary is defined by the path integral on a finite strip, which has only one real modulus, $\beta$ --- choosing a marked point on one of the two boundaries introduces a second real modulus, which is the angular potential when the two ends of the strip are glued into a cylinder.}

The wedge that calculates the wavefunction can also be cut into smaller wedges. This has been described (in somewhat different language) in \cite{Takayanagi:2019tvn,Dong:2019piw}. The total action in this case is not just the sum of the wedge actions. From the wavefunction \eqref{introWF} we see that the gluing relation on-shell is 
\begin{align}\label{glue2wedges}
\Psi_{\tau_1 + \tau_2}(E,J_I) &= e^{-S(E,J_I)/2} \Psi_{\tau_1}(E,J_I) \Psi_{\tau_2}(E,J_I) \ . 
\end{align}
Geometrically,
\begin{align}
	\begin{overpic}[scale=0.8]{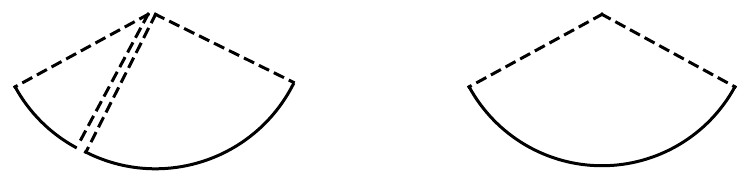}
	\put(37,13){\parbox{0.2\linewidth}{
			\begin{equation*}
				 \text{$=~e^{-S/2}$}
	\end{equation*}}}
\end{overpic}  ~.
\end{align}
The prefactor in \eqref{glue2wedges} comes from the Hayward corner term in the gravitational action \cite{Takayanagi:2019tvn}. The on-shell action of a single wedge has a Hayward term contributing $e^{(1- \psi/\pi)S/2}$ to the wavefunction, where $\psi$ is the internal angle at the corner and $S = A/(4G)$ is the entropy. When $n$ wedges are glued together to make a larger wedge, the extra corner terms must be removed, and this leads to the on-shell relation
\begin{align}
\Psi_{\tau_1 + \tau_2+\cdots + \tau_n}(E,J_I) &= e^{-(n-1)S(E,J_I)/2} \Psi_{\tau_1}(E,J_I) \Psi_{\tau_2}(E,J_I)  \cdots  \Psi_{\tau_n}(E,J_I) \ . 
\end{align}
By gluing the last two edges together to make a disk, we can also form an eternal black hole from $n$ wedges, and the on-shell gravitational action is
\begin{align}
Z(\tau_1+\tau_2+\cdots\tau_n) &= \int dE dJ_I \, e^{-\left(\frac{n}{2}-1\right)S(E,J_I)} \Psi_{\tau_1}(E,J_I) \Psi_{\tau_2}(E,J_I) \cdots \Psi_{\tau_n}(E,J_I) \ . 
\end{align}

\subsubsection*{Outline}
In section \ref{s:setup}, we establish notation and setup the double foliation that is used to specify boundary conditions on internal boundaries. Fixed-$E$ states are considered in section \ref{s:fixedE} and fixed-$(E,J_I)$ states in section \ref{s:fixedEJ}. In each case, we  specify the boundary conditions that define the state $|E\rangle$ or $|E,J_I\rangle$, and calculate the Hartle-Hawking wavefunction $\Psi_{\beta/2}(E)$ or $\Psi_{\beta/2}(E,J_I)$. For fixed-$E$ states, we also describe the dimensional reduction from 3d to 2d gravity in section \ref{ss:jtreduction}, and show that our boundary conditions become those of Harlow and Jafferis \cite{Harlow:2018tqv} when expressed in terms of the dilaton. For fixed $(E,J_I)$ we work out the example of the rotating BTZ black hole in section \ref{ss:BTZ}.

\section{Double foliation}\label{s:setup}

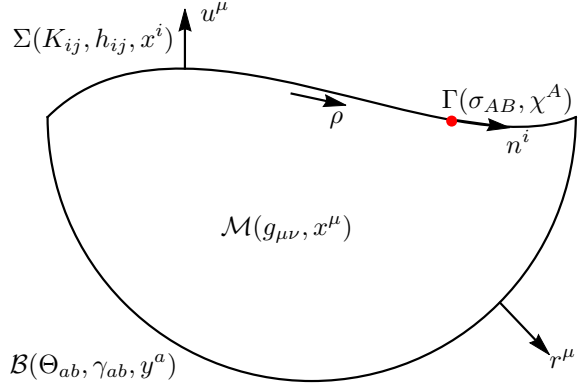
\begin{figure}
\begin{center}
	\begin{overpic}[scale=0.8]{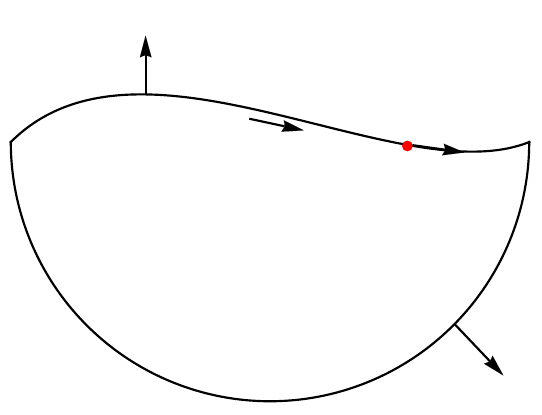}
		\put(75,5){\parbox{0.2\linewidth}{
				\begin{equation*}
					r^\mu
		\end{equation*}}}
		\put(25,30){\parbox{0.2\linewidth}{
				\begin{equation*}
					\mathcal{M}(g_{\mu\nu},x^\mu)
		\end{equation*}}}
		\put(12,68){\parbox{0.2\linewidth}{
				\begin{equation*}
					u^\mu
		\end{equation*}}}
		\put(-10,64){\parbox{0.2\linewidth}{
				\begin{equation*}
					\Sigma(K_{ij},h_{ij},x^i)
		\end{equation*}}}
		\put(-10,5){\parbox{0.2\linewidth}{
				\begin{equation*}
					\mathcal{B}(\Theta_{ab},\gamma_{ab},y^a)
		\end{equation*}}}
		\put(65,53){\parbox{0.2\linewidth}{
				\begin{equation*}
					\Gamma(\sigma_{AB},\chi^A)
		\end{equation*}}}
		\put(67,44){\parbox{0.2\linewidth}{
				\begin{equation*}
					n^i
		\end{equation*}}}
		\put(34,49){\parbox{0.2\linewidth}{
				\begin{equation*}
					\rho
		\end{equation*}}}
	\end{overpic}  
\end{center}
\caption{\small Notation. The spacetime manifold ${\cal M}$ is $D$-dimensional. The $(D-1)$-dimensional hypersurfaces ${\cal B}$ and $\Sigma$ are the asymptotic boundary and internal boundary, respectively. $\Sigma$ is foliated by $(D-2)$-dimensional surfaces $\Gamma$ which correspond to the angular directions in the black hole solution. }\label{fig:notation}
\end{figure}

In $D=d+1$ bulk dimensions, consider an asymptotically-AdS Euclidean spacetime $\M$ with asymptotic boundary $\B$ and internal boundary $\Sigma$, as in figure \ref{fig:notation}. The manifold $\M$ has coordinates $x^\mu$ and metric $g_{\mu\nu}$. The asymptotic boundary $\B$ has coordinates $y^a$, metric $\gamma_{ab}$, extrinsic curvature $\Theta_{ab}$, and outward-pointing unit normal $r^\mu$. The internal boundary $\Sigma$ has coordinates $x^i$, metric $h_{ij}$, extrinsic curvature $K_{ij}$, and outward-pointing unit normal $u^\mu$. 

At asymptotic infinity $\B$, we impose standard (Dirichlet) AdS boundary conditions,
\begin{align}\label{adsbc}
\gamma_{ab} dy^a dy^b = R_c^2 (d\tau^2+ d\Omega_{d-1}^2) + {\cal O}(1)
\end{align}
where $d\Omega_{d-1}^2$ is the metric of the unit $(d-1)$-sphere and $R_c \to \infty$ is the AdS cutoff.  The Hartle-Hawking state is defined by restricting the range of Euclidean time to the finite interval $\tau \in [0, \frac{\beta}{2}]$. 

The canonical momentum conjugate to $\gamma_{ab}$ on $\B$ is
\begin{align}
\pi^{ab} = \frac{\sqrt{\gamma}}{16\pi G} (\Theta^{ab} - \gamma^{ab} \Theta ) \ ,
\end{align}
and the momentum conjugate to $h_{ij}$ on $\Sigma$ is
\begin{align}
P^{ij} = \frac{\sqrt{h}}{16\pi G}(K^{ij} - h^{ij} K) \ .
\end{align}
Foliate $\Sigma$ by hypersurfaces $\Gamma$ labeled by a scalar function $\rho$. The ADM decomposition on $\Sigma$ adapted to this foliation is 
\begin{align}\label{admSigma}
h_{ij}dx^i dx^j = N^2 d\rho^2 + \sigma_{AB} (d\chi^A + N^A d\rho)(d\chi^B + N^B d\rho) \ ,
\end{align}
with $\sigma_{AB}$ as the induced metric on $\Gamma$. We refer to the surface $\Gamma$, which is codimension-2  in $\M$, as the transverse space, and $\rho$ as the `radial' direction. The unit normal to $\Gamma \subset \Sigma$ is $n_i = N \p_i \rho$, and the shift vector is $N^A =-N n^A$ where $n^A$ denotes the transverse components of the vector $n^i$ in the coordinates $x^i = (\rho, \chi^A)$.

Projection tensors onto $\B$, $\Sigma$, and $\Gamma$ are
\begin{align}
\gamma_{\mu\nu} = g_{\mu\nu} - r_\mu r_\nu \ , \qquad
h_{\mu\nu} = g_{\mu\nu} - u_\mu u_\nu \ , \qquad
\sigma_{ij} = h_{ij} - n_i n_j  \ ,
\end{align}
and we define
\begin{align}
\sigma^i_A = \frac{\p x^i}{\p \chi^A} \ .
\end{align}
These satisfy completeness relations such as $\sigma^{ij} = \sigma^{AB} \sigma^i_A \sigma^j_B$ which can be used to decompose tensors into normal and tangential parts.

The normal vector fields $u^\mu$ and $n^\mu$ are \textit{a priori} defined only on $\Sigma$. It is convenient to extend them to a neighborhood of $\Sigma$, which is done arbitrarily subject to the conditions $u^2=n^2=1$ and $u\cdot n = 0$. This also extends the definition of the transverse space $\Gamma$ to a neighborhood of $\Sigma$, with the corresponding projector
\begin{align}\label{sigmaUV}
\sigma_{\mu\nu} = g_{\mu\nu} - n_\mu n_\nu - u_\mu u_\nu \ .
\end{align}

Under the double foliation, the momentum density on the surface $\Sigma$, given by $j^i = - 2 P^{ij}n_j/\sqrt{h}$, decomposes into normal and tangental components 
\begin{align}
j^i &=  qn^i + \sigma_A^i j^A  \ , 
\end{align}
with
\begin{align}\label{defq}
q &= - \frac{2 P_{ij} n^i n^j}{\sqrt{h}} = \frac{1}{8\pi G}K^{ij}\sigma_{ij} \\
j_A &= - \frac{2\sigma^i_A P_{ij}n^j}{\sqrt{h}} = -\frac{1}{8\pi G} \sigma^i_A K_{ij} n^j \ . 
\end{align}
The quantity $j_A$ is interpreted as the momentum density of the codimension two surface $\Gamma$. In the black hole solution, the index $A$ runs over all of the angular directions, whereas $I = 1\dots \lfloor \frac{D-1}{2} \rfloor$ is used to denote the angular directions corresponding to commuting $U(1)$ isometries generated by $\sigma_I^i$, with conserved angular momenta $J_I$.

The normal component $q$ can be related to the transverse area element as
\begin{align}\label{qArea}
q &= \frac{1}{8\pi G} \frac{1}{\sqrt{\sigma}} u^\mu \p_\mu \sqrt{\sigma} \ .
\end{align}
Thus $q$ is the expansion of the surface $\Gamma$ in the direction $u$. In 2d gravity, the role of the transverse area is played by the dilaton, so upon dimensional reduction $q$ is proportional to the normal derivative of the dilaton.

The quantity $q$ is the quasilocal energy of Brown and York \cite{Brown:1992bq,Brown:1992br,Brown:2000dz}. However, it is the quasilocal energy on the internal boundary $\Sigma$, not at infinity, so it is not related to the energy of the black hole on shell.

\textit{Derivation of \eqref{qArea}:}
First note $\sigma^{ij} K_{ij} = \sigma^{\mu\nu}K_{\mu\nu} = \frac{1}{2} \sigma^{\mu\nu} {\cal L}_u g_{\mu\nu} = \frac{1}{2}\sigma^{\mu\nu}  {\cal L}_u \sigma_{\mu\nu}$, where in the last equality we used $n^2=1$ and $u\cdot n = 0$. The transverse area element $\sigma_{AB}$ satisfies $\delta \log \sqrt{\sigma} = \frac{1}{2} \tr \sigma^{-1} \delta \sigma$, so $u^\mu \p_\mu \log \sqrt{\sigma} = \frac{1}{2} \sigma^{AB} {\cal L}_u \sigma_{AB}$. Using  ${\cal L}_u \sigma_{AB} = \sigma_A^\mu \sigma_B^\nu {\cal L}_u \sigma_{\mu\nu}$ this gives $u^\mu \p_\mu \log \sqrt{\sigma} = \frac{1}{2}\sigma^{\mu\nu} {\cal L}_u \sigma_{\mu\nu}$, and \eqref{qArea} follows.

\section{Fixed-$E$ states}\label{s:fixedE}

We will first discuss a bulk state $|E\rangle$ with fixed energy. This state is defined by imposing boundary conditions on the internal boundary $\Sigma$. Our task is to define appropriate boundary conditions, find the saddle corresponding to the overlap $\langle E| \Psi_{\beta/2}\rangle$, and calculate its on-shell action to determine the wavefunction.

\subsection{Boundary condition}

On the internal boundary $\Sigma$, we impose the mixed boundary conditions
\begin{align}\label{fixedEbc}
q &= 0  \\
N^A &= 0 \notag\\
\sigma_{AB} d\chi^A d\chi^B &= r_+(E)^2 \cosh^2 \rho d\Omega_{d-1}^2 \ , \notag
\end{align}
where $\rho \in ( - \rho_c, \rho_c)$, with $\rho_c = \log \frac{2R_c}{r_+(E)}$. The internal boundary $\Sigma$ meets the asymptotic AdS boundary at $\rho = \pm \rho_c$. For $d>2$, the function $r_+(E)$ is defined implicitly by
\begin{align}
E = \frac{(d-1) \Omega_{d-1} r_+^{d-2} }{16\pi G}(1+r_+^2) \ ,
\end{align}
where $\Omega_{d-1} = \frac{2\pi^{d/2}}{\Gamma(d/2)}$ is the area of the unit $(d-1)$-sphere. We have chosen $\sigma_{AB}$ to match the spatial geometry of an eternal black hole with energy $E$. In $d=2$, there is a shift by the Casimir energy, so the relation is $E = \frac{r_+^2}{8G}$.

The particular choice of radial profile $\cosh^2\!\rho$ in \eqref{fixedEbc} is mostly arbitrary. Since we have not fixed the lapse $N$, different choices here will lead to black hole solutions in different coordinates. What is important is that $\cosh^2\rho$ diverges near the boundaries, and the codimension-2 surface $\Gamma$ has minimal area
\begin{align}
A(E) &= r_+^{d-1}(E)\Omega_{d-1} \ .
\end{align}

\subsection{Variational principle}
The gravitational action must be compatible with the boundary condition such that there is well defined variational principle. For Dirichlet boundary conditions, fixing the induced metric on both boundaries $\B$ and $\Sigma$, the appropriate action is
\begin{align}\label{fixedEaction}
I &= -\frac{1}{16\pi G} \int_{\cal M} d^{d+1}x \sqrt{g}(R -2\Lambda) - \frac{1}{8\pi G} \int_{\cal B} d^dy \sqrt{\gamma}\Theta - \frac{1}{8\pi G} \int_{\Sigma} d^d x \sqrt{h}K + I_{\rm ct} \ .
\end{align}
There is a GHY boundary term at both the asymptotic boundary ${\cal B}$ and the internal boundary $\Sigma$, and $I_{\rm ct}$ is the usual counter-term added at the asymptotic boundary to renormalize the gravitational action \cite{Balasubramanian:1999re}.
 There is no counterterm at the internal boundary $\Sigma$.

In fact, the same action \eqref{fixedEaction} also works for the fixed-$E$ boundary conditions in \eqref{fixedEbc} ---  we do not need to change the boundary terms. The on-shell variation is
\begin{align}
\delta I &= \frac{1}{2}\int_{\B} d^d y \sqrt{\gamma}T^{ab} \delta \gamma_{ab}  + \int_{\Sigma} d^d x P^{ij} \delta h_{ij} \ , 
\end{align}
where  $T^{ab}$ is the Brown-York stress tensor at the asymptotic boundary \cite{Brown:1992br, Balasubramanian:1999re},
\begin{align}
T^{ab} &= \frac{1}{8 \pi G}\left(\Theta^{ab} - \gamma^{ab} \Theta\right) + T^{ab}_{\rm ct} \ .
\end{align}
Adopting the decomposition in \eqref{admSigma}, the variation of $h_{ij}$ is\footnote{\textit{Derivation of \eqref{eq:hij_decompose}:} 
Following \cite{Brown:1992br,Brown:2000dz,Brown:1992bq}, the quantity $\sigma^A_i$ is defined as the transverse components of the tensor $\sigma^j_i = \delta^j_i - n^j n_i$ in the ADM coordinates $x^i = (\rho, \chi^A)$. Using $N^A = -N n^A$, we have
\begin{align}
\sigma^A_idx^i = d\chi^A + N^A d\rho \ , \qquad  \ .
\end{align}
Vary \eqref{admSigma},
\begin{align}
\delta h_{ij} dx^i dx^j &= 2N \delta N d\rho^2 + 2 \sigma_{AB} \delta N^A d\rho(d\chi^B + N^B d\rho) + 
\delta \sigma_{AB}(d\chi^A + N^A d\rho)(d\chi^B + N^B d\rho)\notag \\
&= 2N\delta N d\rho^2 + 2\sigma_{AB}\sigma^B_i \delta N^A d\rho  dx^i + \delta \sigma_{AB} \sigma^A_i \sigma^B_j dx^i dx^j
\end{align}
and using $n_i dx^i = N d\rho$ gives \eqref{eq:hij_decompose}.}
\begin{align}\label{eq:hij_decompose}
		\delta h_{ij} &= \frac{2}{N}n_i n_j \delta N + \frac{2}{N}\sigma_{A(i}n_{j)} \delta N^A +\sigma^A_{i}\sigma^B_{j}\delta \sigma_{AB}~.
\end{align}
We obtain the boundary term on $\Sigma$
\begin{align}\label{eq:bdy_Sigma}
		 \int_{\Sigma} d^{d} x P^{ij} \delta h_{ij} &=   \int_{\Sigma} d^{d} x \sqrt{\sigma}\left(- q\delta N-j_A \delta N^A+ \frac{1}{\sqrt{\sigma}}P^{ij}\sigma^A_{i}\sigma^B_{j}\delta \sigma_{AB}\right)~.
\end{align}
The fixed-$E$ boundary condition \eqref{fixedEbc} has $q=0$ and $N^A, \sigma_{AB}$ fixed, so the variation is zero on shell. This establishes that this action is compatible with the fixed-$E$ boundary condition.

\subsection{Wavefunction}\label{s:wavefunction}
The semiclassical wavefunction is 
\begin{align}
\Psi_{\beta/2}(E) &= e^{-I} \ , 
\end{align}
where $I$ is the on-shell action of the solution satisfying the boundary conditions \eqref{adsbc} on $\B$ and \eqref{fixedEbc} on $\Sigma$. For fixed-$E$ boundary conditions, the corresponding solution is a wedge of the eternal black hole of energy $E$, or `pacman', with the opening angle of the wedge determined by $\beta$. See figure \ref{fig:wavefunction-pacman}. This is identical to the situation in JT gravity \cite{Harlow:2018tqv}, but now in higher dimensions. The metric is
\begin{align}\label{ebh}
ds^2 = f(\rho)d\tau^2 + \frac{g(\rho)}{f(\rho)} d\rho^2 + r_+(E)^2 \cosh^2 \rho d\Omega_{d-1}^2 \ , 
\end{align}
where
\begin{align}
f = 1 + r_+(E)^2 \cosh^2 \rho - (1+r_+(E)^2)\cosh^{2-d} \rho , \quad
g = r_+(E)^2 \sinh^2 \rho \ .
\end{align}
The wedge geometry has this same metric but with the coordinate range $\tau \in (0, \frac{\beta}{2})$, as required by the boundary conditions. The black hole with energy $E$ has inverse temperature
\begin{align}
\beta(E) = \frac{4\pi r_+(E)}{d r_+(E)^2 + d-2 } \ .
\end{align}
This is not necessarily related to $\beta$, which is an independent parameter --- thus the classical solution used to calculate $\Psi_{\beta/2}(E)$ is  a wedge covering a fraction $\frac{\beta/2}{\beta(E)}$ of the eternal black hole at energy $E$.

Let's check that this solution satisfies the boundary conditions. The induced metric on the internal boundary is
\begin{align}
ds_{\Sigma}^2 = h_{ij}dx^i dx^j = \frac{g(\rho)}{f(\rho)}d\rho^2 + r_+(E)^2 \cosh^2 \rho d\Omega_{d-1}^2 \ .
\end{align}
This manifestly satisfies the boundary conditions on $N^A$ and $\sigma_{AB}$ specified in \eqref{fixedEbc}. The last boundary condition to check is $q = 0$. Away from the corner, this clearly holds, because a fixed-$\tau$ slice of the eternal black hole has $K_{ij} = 0$. At the corner, we use the formula \eqref{qArea}, which states that $q$ is proportional to the normal derivative of $\log \sqrt{\sigma}$. This vanishes at the corner because the corner lies at the Euclidean horizon, which is a surface of extremal area. (In other words, if the corner is regulated by replacing it with a smooth surface in a small neighborhood of the horizon, then $q=0$ on this surface, independent of the regulator.) Thus, all the boundary conditions is satisfied by the extremal area, and we can proceed to calculate the on-shell action.

The on-shell action of the (entire) eternal black hole is
\begin{align}\label{ibhe}
I(E) = \beta(E)  E - S(E) \ , 
\end{align}
where $S(E) = \frac{1}{4G}\mbox{Area}$ is the entropy. 
The on-shell action of the wedge, using the action \eqref{fixedEaction} appropriate to fixed-$E$ boundary conditions, is 
\begin{align}\label{iwedge1}
I = \frac{\beta}{2\beta(E)} I(E)  - \frac{1}{8\pi G}\int_\Sigma d^d x \sqrt{h} K \ ,
\end{align}
where we have used the fact that the wedge covers a fraction $\frac{\beta}{2\beta(E)}$ of the full black hole. The second term is the contribution of the internal boundary. Away from the corner, $\Sigma$ is a fixed-$\tau$ slice of the eternal black hole, with $K=0$, but there is a finite corner contribution \cite{Hayward:1993my}. The corner term, which can be calculated by applying the Gauss-Bonnet theorem, contributes
\begin{align}
\left. \int_{\Sigma} d^dx \sqrt{h}K \right|_{\rm corner}  &= (\pi - \psi) \int_\Gamma \sqrt{\sigma}
\end{align}
where $\psi$ is the interior angle at the corner. (There are no corner contributions from the joints at the asymptotic boundary, because the angles add up to $\pi$.) The wedge has $\psi = \pi \beta/\beta(E)$, so using \eqref{ibhe}-\eqref{iwedge1} we obtain
\begin{align}\label{iestate}
I= -\frac{1}{2}S(E) + \frac{\beta}{2}E \ .
\end{align}
The semiclassical wavefunction is therefore
\begin{align}
\Psi_{\beta/2}(E) = \exp\left( \frac{1}{2}S(E)  - \frac{\beta}{2}E \right)~.
\end{align}

\subsection{Reduction to JT gravity}\label{ss:jtreduction}
Two-dimensional Jackiw-Teitelboim gravity can be obtained by dimensional reduction in (at least) two different ways. It is the effective theory of near-extremal black holes \cite{Almheiri:2014cka,Mertens:2018fds,Ghosh:2019rcj}, and it also arises as the spherically symmetric sector of 3d gravity \cite{Gross:2019ach}. We will use the latter. The goal is to show that the reduction of the state $|E\rangle$ in three dimension is the state $|E\rangle_{JT}$ defined in \cite{Harlow:2018tqv}. 

Following \cite{Gross:2019ach}, we start in three dimensions with the action \eqref{fixedEaction}, and counterterm \cite{Balasubramanian:1999re}
\begin{align}
I_{\rm ct}^{(3d)} &= \frac{1}{8\pi G}\int_{\B} d^2 y \sqrt{\gamma} \ .
\end{align} 
Assuming a $U(1)$ isometry in the transverse direction, the ansatz for the 3d spacetime metric is
\begin{align}
ds^2_{\M} &= g^{(2)}_{ab} dw^a dw^b + \Phi^2(w^a) d\phi^2 \ , 
\end{align}
with $\phi \sim \phi + 2\pi$. (We are temporarily using $a,b,c,\dots$ for 2d spacetime indices in this subsection; elsewhere these are indices on $\p \M$, but we've run out of letters.) With this ansatz, the intrinsic and extrinsic curvatures are
\begin{align}
R = R^{(2)} - 2 \Phi^{-1} \p^2 \Phi , \quad
\Theta = \Theta^{(1)} + \Phi^{-1}r^\alpha  \p_\alpha \Phi , \quad
K &= K^{(1)}  + \Phi^{-1} u^\alpha \p_\alpha \Phi \ ,
\end{align}
where $R^{(2)}$ is the scalar curvature of the two-manifold $\M_2$ and $K^{(1)}$, $\Theta^{(1)}$ are the extrinsic curvatures at the boundaries of the 2d manifold. Plugging into the action gives
\begin{align}
I_{\rm red} &=
 -\frac{1}{16 \pi G_2} \int_{{\cal M}_2}  \sqrt{g^{(2)}} \Phi(R^{(2)}+2) - \frac{1}{8\pi G_2} \int_{\B_2}  \sqrt{\gamma^{(1)}}\Phi (\Theta^{(1)}-1) - \frac{1}{8\pi G_2} \int_{\Sigma_2} \sqrt{h^{(1)}}\Phi K^{(1)} 
\end{align}
with $2\pi G_2 = G$. 

Having reviewed the dimensional reduction of 3d gravity to JT gravity, we now consider the boundary conditions \eqref{fixedEbc} that define the bulk state $|E\rangle$. Using $K^{ij}n_in_j=K^{(1)}$, the reduction of $q$ defined in \eqref{defq} is
\begin{align}
q = \frac{1}{8\pi G} \Phi^{-1} u^\alpha \p_\alpha \Phi \ .
\end{align}
We have assumed $N^A=0$ in the reduction ansatz, so the remaining boundary conditions are
\begin{align}
\sigma_{\phi\phi} &= \Phi^2 = r_+(E)^2 \cosh^2 \rho \\
u^\alpha \p_\alpha \Phi &= 0 \ .
\end{align}
These are the same boundary conditions used in \cite{Harlow:2018tqv} to define the state $|E\rangle_{JT}$ in JT gravity, as claimed.

\section{Fixed-$(E,J)$ states}\label{s:fixedEJ}

\subsection{Boundary conditions and action}\label{ss:ejbc}

Now we will define a bulk state $|E, J_I\rangle$ labeled by energy and angular momentum. This involves fixing $j_A$ on the internal boundary $\Sigma$. The action \eqref{fixedEaction} is not stationary in this case, as can be seen from the variation \eqref{eq:bdy_Sigma}, so we need to add a boundary term.  

Let $\rho$ be the scalar field that defines the foliation of $\Sigma$ by $\Gamma$, and $N^i = \sigma^i_A N^A$ the shift vector as in \eqref{admSigma}. The action that we will use for fixed-$(E,J_I)$ states is
\begin{align}
\tI &= I - \frac{1}{8\pi G} \int_{\Sigma} d^dx \sqrt{h} N_i K^{ij} \p_j \rho \ ,
\end{align}
with $I$ given in \eqref{fixedEaction}. At the expense of losing manifest covariance, this can also be rewritten using the definitions in section \ref{s:setup} as
\begin{align}
\tI &= I + \int_{\Sigma} d^d x \sqrt{\sigma} j_A N^A \ . 
\end{align}
Using the variation of $I$ from \eqref{eq:bdy_Sigma}, the boundary term at $\Sigma$ in the on-shell variation is now
\begin{align}
 \delta \tI \, \big|_{\Sigma}  &= \int_\Sigma d^d x  \left( -q\sqrt{\sigma} \delta N +  N^A \delta(j_A \sqrt{\sigma}) + P^{ij} \sigma_i^A \sigma_j^B \delta \sigma_{AB} \right)   \ .
\end{align}
This allows us to impose boundary conditions with $q=0$ and fixing $j_A$ and $\sigma_{AB}$. The boundary conditions that define the state $|E, J_I\rangle$ are
\begin{align}\label{fixedEJstate}
q &= 0 \\
j_A &= \bar{j}_A  \notag \\
\sigma_{AB} &= \bar{\sigma}_{AB} \ , \notag
\end{align}
where $\bar{j}_A$ and $\bar{\sigma}_{AB}$ are the momentum density and transverse metric of the rotating black hole solution with energy $E$ and angular momenta $J_I$. Recall that the index $A$ runs over all angular directions, while $I$ runs over the angular directions corresponding to independent angular momenta.

A similar microcanonical action was considered by Brown and York \cite{Brown:1992bq}. If we choose the standard Dirichlet boundary terms at the asymptotic boundary $\B$ and the Brown-York microcanonical boundary terms at the internal boundary $\Sigma$, the action is
\begin{align}
I_{BY} &= 
-\frac{1}{16\pi G} \int_{\cal N} d^{d+1}x \sqrt{g}(R -2\Lambda) - \frac{1}{8\pi G} \int_{\cal B} d^dy \sqrt{\gamma}\Theta + I_{\rm ct} \\
&\qquad  - \frac{1}{8\pi G} \int_{\Sigma} d^d x \sqrt{h}(K^{ij} n_jn_j + N_i K^{ij} \p_j \rho)  \notag \ . 
\end{align}
This is related to the above by
\begin{align}
I_{BY} &= \tI  + \int_{\Sigma}d^dx  \sqrt{\sigma} N q
\end{align}
The on-shell variation of this action at the internal boundary
\begin{align}
\delta I_{BY} \, \big|_{\Sigma} &= 
 \int_\Sigma d^d x  \left(N \delta(q \sqrt{\sigma}) +  N^A \delta(j_A \sqrt{\sigma}) + P^{ij} \sigma_i^A \sigma_j^B \delta \sigma_{AB} \right) \ ,
\end{align}
so this action is appropriate for fixing $q$, $j_A$, and $\sigma_{AB}$. In our case \eqref{fixedEJstate}, we are fixing $q$ to the special value $q=0$, so we are free to use either $\tI$ or $I_{BY}$ --- they both have good variational principles. 
They agree on shell, so this choice does not affect the wavefunction.

Although the boundary terms in the action are the same, the logic we have followed here differs from Brown and York in an important respect. Our microcanonical boundary condition is imposed on the internal boundary, and the quasilocal energy imposed there is always $q = 0$, regardless of the black hole energy. In \cite{Brown:1992bq}, the boundary conditions are imposed on the asymptotic boundary, and $q$ (there called $\varepsilon$) is fixed to the nonzero quasilocal energy density of the black hole. 

\subsection{Wavefunction}
The state $|E, J_I\rangle$ is defined by the boundary condition \eqref{fixedEJstate} at the internal boundary $\Sigma$. Imposing the thermofield double boundary condition \eqref{adsbc} at the asymptotic boundary defines the state $|\Psi_{\beta/2}\rangle$. The wavefunction is the overlap
\begin{align}
\Psi_{\beta/2}(E,J_I) &= \langle E, J_I | \Psi_{\beta/2}\rangle 
\end{align}
which is given semiclassically by $e^{-\tI}$, with $\tI$ the on-shell action of the saddle satisfying these boundary conditions. Similar to the discussion of fixed-$E$ states in section \ref{s:fixedE}, the saddle is a wedge of the eternal black hole at energy $E$ and angular momenta $J_I$. The wedge has $\tau \in (0, \beta/2)$, which covers a fraction $\frac{\beta/2}{\beta(E,J_I)}$ of the eternal black hole; see figure \ref{fig:wavefunction-pacman}. By construction, this solution satisfies all the boundary conditions, including at the corner. The boundary condition $q=0$ forces the corner to occur at an extremal surface.

\subsubsection*{Rotational corner term}
\begin{figure}
	\begin{center}
		\begin{overpic}[scale=0.6]{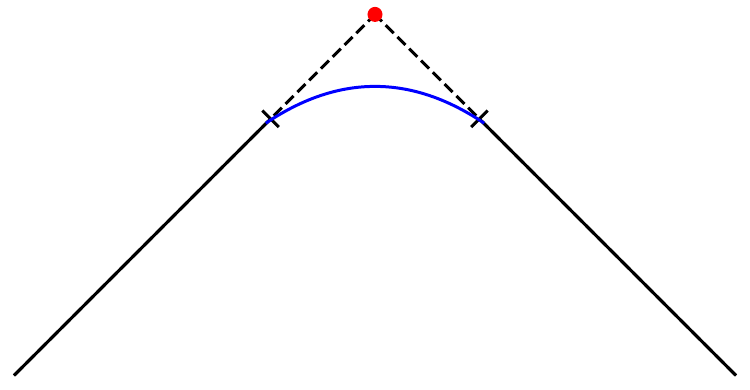}
			\put(30,33){\parbox{0.2\linewidth}{
					\begin{equation*}
					\color{blue}{\mathcal{C}}
			\end{equation*}}}
		\end{overpic}  
	\end{center}
	\caption{\small Regulated corner surface near the Euclidean horizon. }\label{fig:regulatedCorner}
\end{figure}

To calculate the on-shell action we must consider the contribution of the new term, $\int_{\Sigma} \sqrt{\sigma} j_A N^A$. Away from the corner, this term vanishes, because the eternal black hole has $N^A=0$ on surfaces of constant $\tau$. There is, however, a finite corner contribution. There are two ways to calculate it. The first is to view $\Sigma$ as a manifold with two parts, glued together by a twist at the the corner, which introduces a delta function in $N^A$ at the horizon. This method will be demonstrated below for the example of the BTZ black hole. Here we will use the second method, which is to regulate the corner in a small neighborhood of the horizon by replacing it with an arbitrary curve in the $(\tau, r)$ plane, as in figure \ref{fig:regulatedCorner}. On the regulated corner, denoted by ${\cal C}$, the shift vector $N^A$ of the rotating black hole solution has non-vanishing components in the angular directions $\chi^I$ corresponding to conserved angular momenta. Thus, the extra boundary term is
\begin{align}\label{boundarytermC}
I_{\cal C} &:= \int_{\cal C} d^d x \sqrt{\sigma} j_A N^A = \int_{\cal C} d\rho d^{d-1}\chi \sqrt{\sigma} j_I N^I   = \int_{\cal C}d\rho N^I \int_{\Gamma} d^{d-1}\chi \sqrt{\sigma}j_I
\end{align}
where $\rho$ is the coordinate that labels the foliation of ${\cal C}$ as in \eqref{admSigma}. (On ${\cal C}$, this $\rho$ differs from the usual radial coordinate of the black hole, because ${\cal C}$ is not a fixed-$\tau$ surface.) In the last equality we used the fact that the shift is independent of $\chi^A$ in the black hole solution, in the limit where ${\cal C}$ approaches the horizon.

The transverse integral in \eqref{boundarytermC} gives the conserved angular momentum of the black hole,
\begin{align}\label{jjrel}
\int_{\Gamma} d^{d-1}\chi \sqrt{\sigma}j_I &= -i J_I~.
\end{align}
Recall our convention is such that the right-hand side is real for a Euclidean black hole with a real metric. By conservation of the quasilocal stress tensor, \eqref{jjrel} holds on any cross-section $\Gamma$ of the complete boundary, $\Sigma \cup \B$. In particular this integral takes the same value on the interior boundary as it does on the asymptotic AdS boundary, where the angular momentum is usually defined.  (By contrast the quasilocal energy $q$ is not conserved along $\Sigma$, so it is allowed to be zero at the bifurcation surface and nonzero at infinity.)

Now \eqref{boundarytermC} becomes
\begin{align}
I_{\cal C} = -i J_I \int_{\cal C} d\rho N^I \ .
\end{align}
To calculate the $\rho$ integral, let us momentarily consider the entire eternal black hole solution. The Euclidean thermal circle at finite angular potential is $(\tau, \chi^I) \sim (\tau + \beta, \chi^I + \theta^I)$, so integrating around any $S^1$ that circles the Euclidean horizon,
\begin{align}
 \oint d\chi^I = \theta^I
\end{align}
This integral can be re-expressed in terms of the lapse as 
\begin{align}
\oint d\chi^I = - \oint d\rho N^I
\end{align}
Now returning to our calculation fo the wedge action, the regulated corner ${\cal C}$ is a fraction $\frac{\beta}{2\beta(E,J_I)}$ of an $S^1$ circling the horizon, so\footnote{As a check, it is straightforward to confirm this result if we choose the regulator surface ${\cal C}$ to be a small arc with fixed $r$ in the black hole. Then along this arc, $\rho$ (the coordinate labeling the foliation) is equal to $\tau$  (the Euclidean time of the black hole), so $\int_{\cal C}d\rho = \frac{\beta}{2}$ and $N^I =- i \Omega^I(E,J_I) = - \theta^I(E.J_I)/\beta(E,J_I)$.}
\begin{align}
\int_C d\rho N^I = -\frac{\beta}{2\beta(E, J_I)} \theta^I(E,J_I) \ . 
\end{align}
Therefore we find the corner term
\begin{align}\label{ICres}
I_{\cal C}=  i \frac{\beta}{2\beta(E,J_I)} \theta^I(E,J_I) J_I \ .
\end{align}

\subsubsection*{Calculating the wavefunction}
Following the same steps as in section \ref{s:wavefunction} we can now calculate the on-shell action of the wedge as 
\begin{align}
\tI &= \frac{\beta}{2\beta_0} I_0 - \frac{1}{8\pi G} \int_\Sigma d^d x \sqrt{h} K + \tI_{\cal C}\\
&= 
\frac{\beta}{2\beta_0} \left( \beta_0 E - i \theta^I_0 J_I - S(E, J_I)\right) - \frac{1}{8\pi G} \int_\Sigma d^d x \sqrt{h} K + \tI_{\cal C} \notag \\
&= -\frac{1}{2}S(E,J_I) + \frac{\beta}{2}E   \ .  \notag
\end{align}
where $\beta_0 = \beta(E,J_I)$, $\theta^I_0 = \theta^I(E,J_I)$, and the action of the full eternal black hole is $I_0 = I(E,J_I)$.
Therefore, the semiclassical wavefunction $e^{-\tI}$ is
\begin{align}
\Psi_{\beta/2}(E,J_I) &= \exp\left( \frac{1}{2}S(E,J_I) -\frac{\beta}{2}E\right) \ .
\end{align}

\subsection{Example: Rotating BTZ}\label{ss:BTZ}

The Euclidean metric of the rotating BTZ black hole is \cite{Banados:1992wn}
\begin{align}
		ds^2 &= f(\rho) d\tau^2+\frac{g(\rho)}{f(\rho)}d\rho^2+r_+^2 \cosh^2 \rho \left( d\phi + \frac{i r_-}{r_+ \cosh^2 \rho} d\tau \right)^2~,
\end{align}
where 
\begin{equation}
	f(\rho) = r_+^2 \sinh^2 \rho - r_-^2 \tanh^2 \rho~,~~g(\rho) = r_+^2 \sinh^2 \rho~.
\end{equation}
The inner and outer horizons $r_{\pm}$ are functions of the ADM mass $E$ and angular momentum $J$,
\begin{align}
r_\pm^2(E,J)  = 4 G (E \pm \sqrt{E^2 - J^2}) \ . 
\end{align}
and the angular potential is 
\begin{align}
\theta(E,J) = - \frac{ir_-}{r_+} \beta(E,J)\ .
\end{align}
A real Euclidean metric has imaginary $J$, imaginary $r_-$, and real $\theta$. When $J$ is real, the Euclidean metric is complex, but this is not a problem --- it can still be used to calculate a wavefunction in the bulk. The saddlepoint that calculates the wavefunction is the wedge geometry with $\tau \in (-\tfrac{\beta}{4}, \frac{\beta}{4})$, where $\beta$ is an independent parameter.

\begin{figure}
\begin{center}
	\begin{overpic}[scale=0.8]{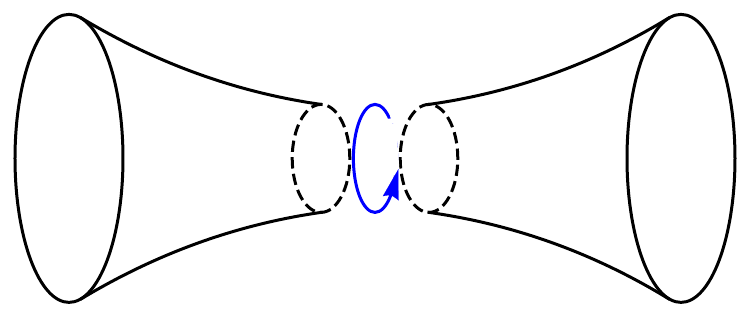}
		\put(35,28){\parbox{0.2\linewidth}{
				\begin{equation*}
					\color{blue}{\alpha}
		\end{equation*}}}
		\put(55,5){\parbox{0.2\linewidth}{
		\begin{equation*}
		\Sigma_+
\end{equation*}}}
		\put(15,5){\parbox{0.2\linewidth}{
		\begin{equation*}
			\Sigma_-
\end{equation*}}}
	\end{overpic}  
\end{center}
\caption{\small On the saddlepoint that calculates the wavefunction of the BTZ black hole, the internal boundary $\Sigma$ consists of two black hole exteriors $\Sigma_-$ and $\Sigma_+$ glued together by a twist $\alpha = \frac{\beta}{2 \beta(E,J)}\theta(E,J)$ at the Euclidean horizon.}\label{fig:Sigmapm}
\end{figure}  

The interior boundary is $\Sigma = \Sigma_- \cup \Sigma_+$, with $\Sigma_{\pm}$ the surfaces at $\tau = \pm \frac{\beta}{4}$. The boundary conditions on $\Sigma$ are
\begin{align}
q = 0 , \quad
j_\phi &= - \frac{i r_-}{8\pi G \cosh\rho}  \ , \quad
\sigma_{\phi\phi}  = r_+^2 \cosh^2 \rho \ . 
\end{align}
On the saddlepoint, the induced metric on $\Sigma$ away from $\rho = 0$ is
\begin{align}
\frac{g(\rho)}{f(\rho)}d\rho^2  + r_+^2 \cosh^2 \rho d\phi^2 \ .
\end{align}
However, the $\phi$ coordinate is not smooth across the joint at the horizon where $\Sigma_+$ meets $\Sigma_-$, so we cannot use this metric globally on $\Sigma$. The angle $\phi$ jumps by an amount $\phi \to \phi + \alpha$  at the corner, with $\alpha = \frac{\theta(E,J)\beta}{2\beta(E,J)}$ as required by the thermal periodicity. This is illustrated in figure \ref{fig:Sigmapm}. The corotating coordinate $\tilde{\phi} = \phi - \frac{\theta(E,J)}{\beta(E,J)}\tau$ is continuous at $\rho=0$, so on $\Sigma$ we can introduce the smooth coordinate $\tilde{\phi} = \phi - \theta(\rho)\alpha$ where $\theta(\rho)$ is the step function. Therefore the induced metric is
\begin{align}\label{sigmaTwist}
ds^2_{\Sigma} &=\frac{g(\rho)}{f(\rho)}d\rho^2 + r_+^2 \cosh^2 \rho \tilde{\phi}^2\\
&= \frac{g(\rho)}{f(\rho)}d\rho^2 + r_+^2 \cosh^2 \rho (d\phi  -\alpha \delta(\rho) d\rho)^2 \ . 
\end{align}
The shift is $N^\phi = -\alpha \delta(\rho)$, and the angular momentum corner term in the on-shell action is
\begin{align}
\int_{\Sigma} \sqrt{\sigma} j_\phi N^\phi &= i \frac{r_+r_-}{4G}\alpha = i \alpha  J 
\end{align}
in agreement with the general result in \eqref{ICres}. The other terms in the on-shell action are straightforward to calculate, leading to
\begin{align}
\tI &= -\frac{1}{2}S(E,J) + \frac{\beta}{2}E \ , 
\end{align}
where $S = \frac{\pi r_+}{2G}$ is the entropy.

\ \\ 

\ \\

\noindent \textbf{Acknowledgments}\\
We thank Jeevan Chandra, Scott Collier, Xi Dong, Yikun Jiang, David Kolchmeyer, Ho Tat Lam, Don Marolf, and Baur Mukhametzhanov for helpful discussions, and Yikun Jiang for collaboration in the early stages of this project. This work is supported by NSF grant PHY-2014071.

\small
\bibliographystyle{ourbst}
\bibliography{gravity.bib}

\end{document}